\documentstyle[12pt]{article}

\newcommand{\sect}[1]{\setcounter{equation}{0}\section{#1}}

\def\be{\begin{equation}}
\def\ee{\end{equation}}
\def\ba{\begin{eqnarray}\samepage}
\def\ea{\end{eqnarray}}

\font\twelvemsa=msam10 scaled 1200

\font\sevenmsa=msam7
\font\fivemsa=msam5
\newfam\msafam
\textfont\msafam=\twelvemsa
\scriptfont\msafam=\sevenmsa
\scriptscriptfont\msafam=\fivemsa
\def\msa{\ifcase\msafam 0\or1\or2\or3\or4\or5\or6\or7\or8\or9\or A\or B\or
C\or D\or E\or F\fi}

\font\twelvemsb=msbm10 scaled 1200

\font\sevenmsb=msbm7
\font\fivemsb=msbm5
\newfam\msbfam
\textfont\msbfam=\twelvemsb
\scriptfont\msbfam=\sevenmsb
\scriptscriptfont\msbfam=\fivemsb
\def\msb{\ifcase\msbfam 0\or1\or2\or3\or4\or5\or6\or7\or8\or9\or A\or B\or
C\or D\or E\or F\fi}

\font\twelveeuf=eufm10 scaled 1200

\font\seveneuf=eufm7
\font\fiveeuf=eufm5
\newfam\euffam
\textfont\euffam=\twelveeuf
\scriptfont\euffam=\seveneuf
\scriptscriptfont\euffam=\fiveeuf
\def\euf{\ifcase\euffam 0\or1\or2\or3\or4\or5\or6\or7\or8\or9\or A\or B\or
C\or D\or E\or F\fi}

\mathchardef\gapprox"3\msa26
\mathchardef\lapprox"3\msa2E

\begin{document}
\title{Soliton solution in dilaton - Maxwell gravity}

\author{Maria Yurova\\
\\Institute of
Nuclear Physics\\Moscow State University
\\Vorobjovy Gory\\119899 Moscow\\Russia
\\e-mail:\,\,yurova@monet.npi.msu.su}

\maketitle


\begin{abstract}
\noindent
The inverse scattering problem method application to construction
of exact solution for Maxwell dilaton gravity system ia considered.
By use of Belinsky and Zakharov L - A pair the solution of the theory
is constructed. The rotating Kerr - like configuration with NUT -
parameter is obtained.
\end{abstract}

\renewcommand{\thepage}{ }
\pagebreak

\renewcommand{\thepage}{\arabic{page}}
\setcounter{page}{1}


\sect{Model discussed}

Recently much attention has been given to the study of gravity models
appearing in the low energy limit of string theory
\cite {1}--\cite {5}.
These models describe various interacting "matter" fields
coupled to gravity; one of them 
considers the system of interacting gravitational,
abelian vector and scalar fields with the action
\begin{equation}
^4S=\int d^4x |g|^{\frac{1}{2}}\left [ -R+2\left (\partial\phi\right )^2
-e^{-2\phi}F^2\right ],
\end{equation}
where $R$ is the Ricci scalar for the metric $g_{\mu \nu}$,
$(\mu = 0, ...,3)$; $F_{\mu \nu} = \partial _{\mu}A_{\nu} -
\partial _{\nu}A_{\mu}$.
We will discuss the stationary and axisymmetric model when both 
the metric and matter fields depend only on two space coordinates.

The four - dimentional line element can be parametrized according 
to $\cite{6}$ 
\begin{equation}
ds_4^2 = g_{\mu \nu}dx^{\mu}dx^{\nu} = f(dt - \omega _\varphi d\varphi)^2 -
f^{-1}ds_3^2,
\end{equation}
and the three - dimentional one can be taken in the Lewis - Papapetrou 
form
\begin{equation}
ds_3^2 = h_{mn}dx^mdx^n = e^{2\gamma}(d\rho ^2 + dz^2) + \rho ^2d\varphi ^2.
\end{equation}

The system under consideration may be completely described by the Einstein
equations defining the metric function $\gamma$, 
as well as by the set of motion equations for the action
\begin{equation}
^2S = \int d\rho dz \rho L,
\end{equation}
where Lagrangian in term of $\sigma$ - model variables $\cite{7}$
has the form
\begin{eqnarray}
L &=& \frac {1}{2}f^{-2}[(\nabla f)^2 + (\nabla \chi + v\nabla u -
u\nabla v)^2] \nonumber \\
&-& f^{-1}[e^{2\phi}(\nabla u)^2  + e^{-2\phi}(\nabla v)^2]
\nonumber \\
&+& 2(\nabla \phi)^2.
\end{eqnarray}
($\nabla$ is connected with the flat two - metric $\delta_{ab}$.)
Here the magnetic $u$ and rotation $\chi$ potentials are introduced by 
"dualization" in three - dimentional formulation of the stationary theory 
instead of the vector potential $\vec A$ and metric function
$\vec \omega$ respectively:
\begin{equation}
\nabla u = fe^{-2\phi}(\sqrt{2}\nabla \times \vec A + \nabla v \times \vec
\omega) + \kappa \nabla v,
\end{equation}
\begin{equation}
\nabla \chi = u\nabla v - v\nabla u - f^2\nabla \times \vec \omega,
\end{equation}
where $v = \sqrt {2} A_0$, and three--dimensional operator
$\nabla$ is connected with the metric $h_{ij}$.

In this letter we will study the case with $\chi$ = $u$ = 0, i. e. the pure electric 
configurations. It can be described by the action $\cite{8}$
\begin{equation}
^2S = \frac {1}{2}\int d\rho dz \rho Tr(J^P)^2,
\end{equation}
where $J^P = (\nabla P)P^{-1}$ and the symmetric $2 \times 2$
matrix $P$ consists of the "matter" fields:
\begin{eqnarray}
P = \left (\begin{array}{crc}
f - v^2e^{-2\phi} &-& ve^{-2\phi}\\
- ve^{-2\phi} &-& e^{-2\phi}\\
\end{array}\right ),
\end{eqnarray}
The set of Euler - Lagrange equations in the form of matrix equation reads
\begin{equation}
\nabla (\rho J^P) = 0,
\end{equation}
and it is for this equation that we will apply the inverse 
scattering problem method. The corresponding four - metric becomes 
static; its space part is determined by the function $\gamma$
\begin{eqnarray}
\gamma _{,z} &=& \frac {\rho}{2}Tr[J^P_{\rho}J^P_{z}],
\\
\gamma _{,\rho} &=& \frac {\rho}{4}Tr[(J^P_{\rho})^2 - (J^P_{z})^2 ].
\nonumber
\end{eqnarray}

As it is well known, having one solution of the theory one can obtain
 anouther one by applying some kind of symmetry transformation. So
by changing the dilaton sign one can construct the pure magnetic field
configuration with the same metric. Similarly the complex discrete
Bonnor transformation $\cite{9}$ maps the system (1.8) into the one
with the Kerr - like space - time metric with NUT parameter.

Thus in this letter we construct the nontrivial asymptotically flat 
configuration from the trivial fields and space - time metric
by using the inverse scattering problem thecnique, and from this 
with the help of symmetry transformations 
we obtain the systems with the other set of field variables. 

\sect{Inverse scattering problem method}

As it has been shown by Belinsky and Zakharov, the vacuum
stationary axially symmetric gravitational equations may be integrated
by use of
the inverse scattering problem method as proposed in
$\cite{10}$ - $\cite{11}$.
This method allows to obtain the n - soliton configurations from the
flat space - time, and in the case of two - soliton solution    
the Kerr - NUT metric with horizons is constructed .

Let us describe in brief the main aspects of general sheme used in further
consideration $\cite{11}$.

The part of vacuum axially symmetric 
Einstein equations determining the metric of subspace  $(t, \phi)$ reads 
\begin{equation}
\nabla (\rho J^g) = 0, \qquad J^g = (\nabla g)g^{-1},
\end{equation}
($\nabla_{i}$ = ${\partial}_{i}$, $i = \rho, \phi $)
where $g$ must satisfy the condition
\begin{equation}
\det g = -\rho^{2}.
\end{equation}
The integration of (2.1) is associated with the L - A pair
\be
[\partial_{z} - \frac{2 \lambda^{2}}{\lambda^{2} + \rho^{2}}
\partial_{\lambda}\,]\, \psi = \frac{\rho {J^g}_{z} - \lambda {J^g}_{\rho}}
{\lambda^{2} + \rho^{2}} \, \psi, 
\ee
\ba
\nonumber
[\partial_{\rho} + \frac{2 \lambda \rho}{\lambda^{2} + \rho^{2}}
\partial_{\lambda}\,]\, \psi = \frac{\rho {J^g}_{\rho} + \lambda {J^g}_{z}}
{\lambda^{2} + \rho^{2}} \, \psi,
\nonumber
\ea
where $\lambda$ is the spectral complex parameter and the function
$\psi$ = $\psi(\lambda, \rho, z)$. Then the solution of (2.1) for
the symmetric metric matrix $g$ represents as:
\be
g(\rho, z) = \psi (0, \rho, z).
\ee
The soliton solutions for the matrix $g$ correspond to the pole divergense
in the spectral parameter complex plane for the matrix $\psi$; its  
pole trajectories are determined by
\be
\mu_{k} (\rho, z)= w_{k}-z \pm [(w_{k}-z)^{2}+{\rho}^{2}]^{\frac{1}{2}},
\quad w_{k}= {\rm const}
\ee
for each pole $k$.

To satisfy the requirement (2.2), it is useful to note, that when $g$
is the solution of Eq. (2.1), the "physical" function
$g^{ph}$ = $ - \rho (- \det g)^{- \frac{1}{2}}g$ is the solution, too.

Hence, omitting the intermediate calculations, the resulting expression
for the metric tensor n - soliton solution has the form:
\be
{g^{ph}}_{ab}(\rho, z)= -\rho^{-n} (\,\prod _{p=1}^{n}\, \mu_{p})\,
[(g_{0})_{ab} -
\sum _{k,l=1}^{n} \,\,{{\Gamma}_{kl}}^{-1}{\mu_{k}}^{-1}{\mu_{l}}^{-1}
{N^{(k)}}_{a}{N^{(l)}}_{b}],
\ee
where
\be
{N}^{(p)}=g_{0}\,{[{{\psi}_{0}}^{-1}({\mu_{p}},\rho,z)]}^{T}\, {m_{0}}^{(p)},
\quad
{\Gamma}_{kl}={({\rho}^{2}+{\mu_{k}}{\mu_{l}})}^{-1}{N^{(k)}}^{T}\,
{g_{0}}^{-1}\, N^{(l)},
\ee
and the column ${m_{0}}^{(p)}$ consists of arbitrairy constants:
\ba
{m_{0}}^{(p)} = \left (\begin{array}{crc}
{C_{0}}^{(p)}& \\
{C_{1}}^{(p)}&\\
\end{array}\right ),
\ea                 
corresponding to the different charactheristics of the source.


\sect{Exact soliton solution}

Now let us apply the inverse scattering problem method to the construction
of the static axial - symmetric two - soliton solution for the Maxwell -
dilaton gravity model (1.8).

If one takes the matrix Euler - Lagrange motion equation (1.10) and distracts 
from the
physical sence of its components, one can see that its form coinsides 
with (2.1) for Einstein gravity. The important fact is that the matrix
dimention of both this expressions is the same. This gives the reason
to believe that the application of the considered above technique allows
to obtain an exact solution of the discussed model.

Our system does not have a condition like (2.2), but the asymptotic behavior
of the metric and fields raquires
\be
P_{\infty}  = \sigma _{3}.
\ee
To satisfy this we determine the "physical" matrix as 
$P^{ph}$ = $-{(-\det P)}^{-\frac{1}{2}} P$, which is also the solution of 
(1.10).
This leads to the restriction $\det P^{ph}$ = $-1$, in other words for all
solutions it must be $f = e^{2\phi}$ (see (1.9)). This limitation being 
the result of the technique used shows, that the pure nontrivial gravity system or
the Einstein - Maxwell system are not contained in the constructed solution.

The initial values of metric and field variables correspond to a flat
space - time and a trivial dilaton and electric configurations (3.1).
Following \cite{11}, we obtain the solution in the Boyer - Lindkuist
coordinats:
\be
\rho = [{(r-m)}^{2} - \sigma^{2}]^{\frac{1}{2}} \sin{\theta}, \qquad
z - z_{1}=(r-m)\cos{\theta},
\ee
where the new constants $\sigma$ = $\frac{1}{2}(w_{1} - w_{2})$ and
$z_{1}$ = $\frac{1}{2}(w_{1} + w_{2})$ are introduced (see (2.5));
as one can see below, $m$ corresponds to the mass of the source.
In analogy with the pure gravity case we impose the conditions on 
the arbitrary constants (2.8):
\ba
{C_{1}}^{(1)}{C_{0}}^{(2)}-{C_{0}}^{(1)}{C_{1}}^{(2)} = \sigma, \qquad
{C_{1}}^{(1)}{C_{0}}^{(2)}+{C_{0}}^{(1)}{C_{1}}^{(2)} = -m, \\
{C_{1}}^{(1)}{C_{1}}^{(2)}-{C_{0}}^{(1)}{C_{0}}^{(2)} = a, \qquad 
{C_{1}}^{(1)}{C_{1}}^{(2)}+{C_{0}}^{(1)}{C_{0}}^{(2)} = -b,
\ea
where $\sigma^{2}$=$m^{2}-b^{2}+a^{2}$.

If we introduce the notation
\be
\Delta=r^{2}-2mr+b^{2}-a^{2}, \qquad
{\delta}^{2}=r^{2}-{(b-a\cos{\theta})}^{2},
\ee
then the components of matrix $P$ can be presented in the form
\be
f=e^{2\phi}=\frac{\Delta+a^{2} {\sin}^{2}\theta}{{\delta}^{2}},
\qquad v=2\, \frac{-br+m(b-a\cos{\theta})}{{\delta}^{2}}.
\ee

The four - dimentional line element corresponds to diagonal metric
with space part resembling the Kerr one:
\ba
ds_4^2 &=& \frac{\Delta+a^{2} {\sin}^{2}\theta}{{\delta}^{2}}\,dt^{2} -
\frac{{\delta}^{2}}{\Delta+a^{2} {\sin}^{2}\theta}\, \Delta \,
{\sin}^{2}\theta \,d\varphi^2
\nonumber \\
&-& \frac{\Delta+a^{2} {\sin}^{2}\theta}
{{\Delta+ \sigma^{2} {\sin}^{2}\theta}}\,[ \,\frac{{\delta}^{2}}{\Delta}\,dr^2
+\,{\delta}^2 \,d\theta ^2].
\ea
Hence one can see that $m$ is the mass of the source and $b$ is proportional
to the electric charge.

As it was noted before, it is interesting to construct one field configuration
from the other using some symmetry transformation. 
By changing the dilaton sign from the electric solution one can obtain 
the pure magnetic solution with the same metric and  $f$ = $e^{-2\phi}$, 
$u$ = $v$, where $f$ and $v$ are determined  by (3.6).

Another interesting case is connected with the complex discrete
transformation allowing to obtain the rotating system with NUT - parameter.
The Bonnor transformation
\be
v \rightarrow -i \chi
\ee
leads to a complex changing of parameters $a$ and $b$ corresponding now
to a rotating moment and a NUT - parameter of the source.
Given transformation  maps the effective  Lagrangian $L(f, v, \phi)$ into 
the one 
$L(f, \chi)$ =  2$L_{E}$, where $L_{E}$ relates to the vacuum gravity. 
One can see that this Lagrangian appears in the Einstein - Maxwell - 
dilaton theory with pseudoscalar Peccei - Quinn axion field (EMDA) 
$\cite {12}$. (As it is known, the theory (1.1) is not always  a 
subsystem of EMDA) . Thus one can leave the framework of the theory under  
consideration 
(1.1) and obtain the EMDA particular solution that  presents six "moduli" 
field 
expressed in terms of two variables $f$ and $\chi$. One of this possible 
models corresponding to the special solution anzats is considered in 
$\cite{13}$; as an example we would like to present the symplest 
case with $f$ = $ - e^{- 2\phi}$ and axion field $\kappa$ = $\chi$, 
related to the system without vector fields.     

Then the field functions have the form:
\be
f= -e^{-2\phi}=\frac{\Delta-a^{2} {\sin}^{2}\theta}{{\delta}^{2}},
\qquad \chi=\kappa= 2\, \frac{-br+m(b-a\cos{\theta})}{{\delta}^{2}}.
\ee
Using (1.7) we obtain
\be
\omega _\varphi = \frac{2}{\Delta-a^{2} {\sin}^{2}\theta}\,
[b \cos \theta\,\Delta\,-\, a\,{\sin}^{2}\theta\,(mr+b^{2})],
\ee
which defines the metric component $g_{t \varphi}$.
Thus the resulting four - metric becames:
\ba
ds_4^2 &=& \frac{\Delta-a^{2} {\sin}^{2}\theta}{{\delta}^{2}}\,[dt
- \omega _\varphi \,d\varphi\,]^{2} -
\frac{{\delta}^{2}}{\Delta-a^{2} {\sin}^{2}\theta}\, \Delta \,
{\sin}^{2}\theta \,d\varphi^2
\nonumber \\
&-& \frac{\Delta-a^{2} {\sin}^{2}\theta}
{{\Delta+ \sigma^{2} {\sin}^{2}\theta}}\,[ \,\frac{{\delta}^{2}}{\Delta}\,dr^2
+\,{\delta}^2 \,d\theta ^2].
\ea

It is easy to see that this expression is close to the Kerr - NUT
metric. The difference is assotiated with the duplication of  
$L_{E}$; thit leads to the fact that the
metric function $\gamma$ = 2$\gamma_{E}$ (see(1.11)), changing the 
three - dimentional metric (1.3).
As the result one has the dependance of the metric on the coordinate $\theta$ 
in  the spherically symmetric case (when $a$ = $b$ = $0$), 
this also concerns  the metric  (3.7).

At the last time one can note that the equation (1.10) is invariant under the
transformation
\be
P \rightarrow P^{-1}.
\ee
It is easy to prove that the solution arising after applying
of (3.12) coincides with the one constructed above with the additional
replacement $z \rightarrow - z$  and $b \rightarrow - b$.

\section{Discussion}

In this letter by use of the inverse scattering problem method the exact
stationary axially symmetric solution of interacting electric - dilaton
gravity system is constructed. This becomes possible because of the
suitable matrix dimention description of the theory under consideration
and as a result of the motion equations. Actually this fact plays
an important role: so,  the direct application of above mentioned technique
for the case of the Einstein - Maxwell - dilaton - axion gravity
 describing by the four - dimentional matrix does not allow
to obtain the solution with "good" asymptotics for metric and fields.
The technique used  fixes the anzatc not containing the nontrivial
gravity or Maxwell - gravity systems as subsystems. 

This demonstrates the possibility of soliton solution construction
for gravity system with fields described by two - dimentional matrices,
and we would hope in further to generalize this for the case of
arbitrairy degree of friedom number.


\bigskip

\hfil{\large\bf Acknowledgements}\hfil

\medskip

\noindent
This work was supported by RFBR grant ${\rm N^0} 00\, 02\, 17135$
\newpage



\begin{thebibliography}{30}
\bibitem {1}
J. Maharana and J.H. Schwarz, Nucl. Phys. {\bf B390} 3, 1993.
\bibitem {2}
J.H. Schwarz and A. Sen, Nucl. Phys. {\bf B411} 35, 1993.
\bibitem {3}
A. Sen, Int. J. Mod. Phys. {\bf A9} 3707, 1993.
\bibitem {4}
A. Sen, Nucl. Phys. {\bf B434} 179, 1995.
\bibitem {5}
A. Sen, Nucl. Phys. {\bf B447} 62, 1995.
\bibitem {6}
W. Israel and G. A. Wilson, J. Math. Phys. {\bf 13} 865, 1972. 
\bibitem{7}
D.V. Galtsov, A.A. Garcia, O.V. Kechkin,
Class. Quant. Grav. {\bf 12} 2887, 1995.
\bibitem{8}
D.V. Gal'tsov, O.V. Kechkin,  Phys. Lett.
{\bf B361} 52, 1995.
\bibitem{9}
W.M.Bonnor, Zeitschrift fur Phys. {\bf 190} 444, 1966.
\bibitem{10}
V.A. Belinsky, V.E. Zakharov,
Sov. Phys. JETP {\bf 48} 985, 1978.

\bibitem{11}
V.A. Belinsky, V.E. Zakharov,
Sov. Phys. JETP {\bf 50} 1, 1979.

\bibitem{12}
D.V. Gal'tsov, A.A. Garcia, O.V. Kechkin,
J. Math. Phys. {\bf 36} 5023, 1995,
D.V. Gal'tsov,  O.V. Kechkin, Phys. Rev.
{\bf D50} 7394, 1994, 
D.V. Gal'tsov, Phys. Rev. Lett. {\bf 74} 2863 1995.
\bibitem{13}
O.V.Kechkin,
Gen. Rel. Grav. {\bf 31} 1075, 1999.
\end{thebibliography}
\end{document}